\documentstyle[11pt,aaspp4,graphicx,subfigure,fancybox]{article}

\newcommand{\cosec}{\mathop{\rm cosec}\nolimits}

\newcommand{\cd}{\hbox{$\cosec\left|\delta\right|$}}

\begin{document}

\title{SUMSS: A Wide--Field Radio Imaging Survey of the Southern Sky. 
I. Science goals, survey design and instrumentation} 

\centerline{(To appear in \emph{The Astronomical Journal})}

\author{D. C.--J. Bock\footnote{present address: Radio Astronomy Laboratory,
University of California at Berkeley, Berkeley, CA 94720, USA.},
 M. I. Large \& Elaine M. Sadler}

\affil{School of Physics, University of Sydney, NSW\,2006, Australia\\
Electronic mail: dbock@astro.berkeley.edu, large@physics.usyd.edu.au, 
ems@physics.usyd.edu.au }


\vspace{10mm}

\begin{abstract}
The Molonglo Observatory Synthesis Telescope, operating at 843\,MHz with a 5 
square degree field of view, is carrying out a radio imaging survey of the sky 
south of declination $-30^{\circ}$.   This survey (the Sydney University 
Molonglo Sky Survey, or SUMSS) produces images with a resolution of $43'' 
\times 43''$ \cd\ and an rms noise level of $\sim$ 1 mJy beam$^{-1}$. SUMSS is
therefore similar in sensitivity and resolution to the northern NRAO VLA Sky
Survey (NVSS; Condon et al.\ 1998). The survey is progressing at a rate of about
1000 square degrees per year, yielding individual and statistical data for many
thousands of weak radio sources. This paper describes the main characteristics
of the survey, and presents sample images from the first year of observation. 
\end{abstract}


\section{Introduction}

The development of radio telescopes with  resolution better than one minute of
arc, and the increasing sophistication of astronomical image processing
techniques, have made it possible to undertake large--scale, sub--arcminute
surveys presenting the data not just in catalogue form, but also as images. 
Such high--resolution radio imaging surveys are analogous to optical Schmidt
sky surveys, and can play a similarly valuable role in discovering  new and 
interesting individual objects and studying the
Universe on the largest scales. In the Northern Hemisphere, sub--arcminute
radio surveys have been completed, or are well advanced, with the VLA (NVSS:
Condon et al.\ 1998, and FIRST: Becker et al.\ 1995) and Westerbork (WENSS:
Rengelink et al.\ 1997). The Sydney University Molonglo Sky Survey (SUMSS)
described in this paper covers areas of sky not accessible to northern
telescopes, and provides radio spectral index information in the regions of
overlap. 

The Molonglo Radio Observatory has a long history of radio survey work. The
original telescope (the 408\,MHz Molonglo Cross) had a resolution of 2.8$'$.
Between 1966 and 1978 its achievements included an imaging survey of the
Galactic plane (Green 1974), acquisition of the data for the Molonglo Reference
Catalogue of radio sources (MRC: Large et al.\ 1981, Large et al.\ 1991) and
the discovery of 186 pulsars (Manchester et al.\ 1978). The 408\,MHz Molonglo
Cross was closed in 1978 and its east--west reflector was used in a new
instrument, the Molonglo Observatory Synthesis Telescope (MOST), an imaging
radio telescope operating at 843\,MHz. A review of the work of the Molonglo
Cross and its conversion to the MOST was given by Mills (1991) at the symposium
celebrating the 25th anniversary of the Molonglo Observatory. 

After more than a decade of successful observations with, and technical
enhancements of, the MOST, it became clear that it would be feasible and
desirable to increase the telescope's field of view from 1 to 5 square degrees. 
This project, which involved the installation of hundreds of low noise
preamplifiers and associated IF amplifiers and phase shifters, was completed in
March 1997. The upgraded telescope now has the capacity to survey the southern
sky to faint limits at a rate of $\geq $1000 square degrees per year. This
survey rate, combined with the relatively low frequency of operation, good
sensitivity (rms noise level $\sim$ 1 mJy beam$^{-1}$) and stable
configuration, makes the upgraded MOST a particularly powerful tool for
carrying out an all--sky radio imaging survey. 

In this paper we first  describe (\S2) the upgraded MOST, its mode of operation
and some characteristics of the wide--field radio images it produces. We then
explain (\S3) how our new southern radio survey has been designed, and discuss
(\S4) some of the planned follow--up science. A second paper (Cram et al.\
1998, in preparation) will discuss the data processing pipeline. Details of the
closely--related MOST Galactic plane survey ($-10^\circ< b < +10^\circ$) will
be published at a later date. 

\section{The Molonglo Observatory Synthesis Telescope}

The MOST (Mills 1981; Robertson 1991) is an east--west array operating at
843\,MHz.  It comprises two co-linear cylindrical paraboloids (`arms'), each
11.6\,m wide and 778\,m long,  separated by a 15\,m gap. The geometrical
aperture of over 18,000\,m$^{2}$ is the largest of any radio telescope in the
southern hemisphere. 
 
The MOST is essentially an array of mechanically steered antenna elements which
return intermediate frequency signals to a central location for image
processing.  The antenna elements of the MOST  are 44 contiguous 17.\,8m
lengths of each arm (referred to as `bays'). Figure 1 
\begin{figure}[tbp]
\centering\includegraphics[height=8cm]{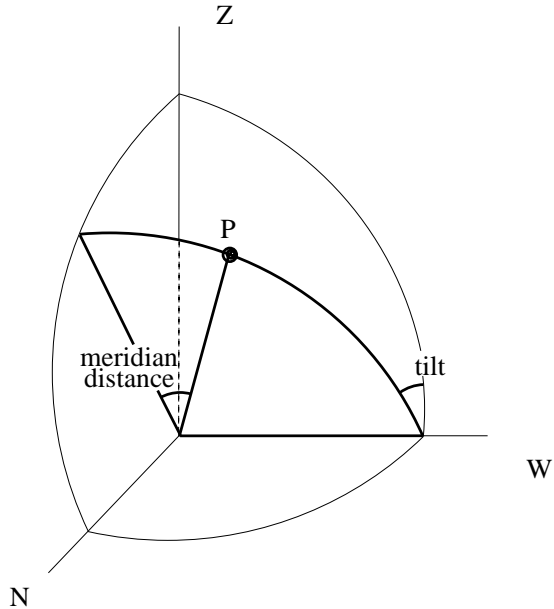}
   \caption{Coordinate system used with the MOST. 
           The position, P, of a MOST beam is described by a right--handed 
           spherical coordinate system, with the pole pointing West. 
           {\em Tilt\ }(analogous to longitude) is measured towards North 
           from the vertical.  {\em Meridian distance\ }(analogous to 
           latitude) is measured towards West from the meridian plane. }
    \label{fig:MOST_schem}
\end{figure}
defines the parameters of
MOST's natural coordinate system (tilt and meridian distance).  Note that the
MOST resembles a conventional alt--az telescope, but with the important
difference that the fixed axis is aligned horizontally (east--west) rather
than vertically.   All 88 bays are steered to the same tilt by rotation of the
reflectors about their east--west axis. Each bay is also steered to the same
meridian distance (angular distance from the meridian plane) by differential
rotation of the ring antennas (spaced at 0.54 $\lambda$) which comprise the
line feed. 

Since the MOST is an east--west array, full aperture synthesis requires
continuous observation for 12 hours.   During an observation, the bays
track the chosen field center, while the array tracking is
achieved by introducing appropriate, frequently updated phases and delays into
the signals from each bay. The meridian distance of the MOST is restricted to
the range  $\pm67^\circ$. Thus, in principle, the MOST can fully synthesize in
a 12--hour observation any field south of declination $-23^\circ$. 
Fields north of $-23^\circ$ can  be only partially synthesised. In practice,
the northern declination limit for full synthesis is about $-30^\circ$. 

Multi-beaming circuits form a comb of 64 hard--wired real--time fan beams at
the Nyquist spacing of 22$''$. MOST images are formed by the technique of
earth--rotation fan beam synthesis (Perley 1979; Crawford 1984). In its
original configuration, the MOST had a basic field size of 23$'$, extendable to
70$'$ by time sharing. In preparation for the present survey (SUMSS), 352 new
low-noise pre-amplifiers and associated  phase shifters have been installed
(Large et al.\ 1994).  Computer control of these phase shifters enables the
time--shared field size to be extended to 160\,arcminutes (possibly more) by a
novel technique, described in the Appendix. 
  
Table 1 summarizes some characteristics of synthesis observations with the MOST
in its wide--field mode. 
\begin{table}
\tablenum{1}
\centering
\begin{tabular}{|l|l|}
\hline
\hline
Parameter &  Value\\
\hline
Frequency    &   843\,MHz\\
Bandwidth    &     3\,MHz\\
Declination range (12\,hr synthesis)& $-90^{\circ}\leq\delta<-30^{\circ}$\\
Maximum field size (RA$\times$Dec) &  Elliptical, $163'\times 163'$\cd\\ 
rms noise level (1$\sigma$) & 1--2\,mJy beam$^{-1}$\\
Resolution (FWHM)  &  $43''\times 43''$\cd\\
Polarization &  Right circular (IEEE)\\
\hline
\end{tabular}
\caption{Key characteristics of the MOST in its wide--field mode. } 
\end{table}

\subsection{Properties of the MOST wide--field images }

\label{sec:wf_im_char}
The radio images produced by MOST differ from those produced by
minimum--redundancy radio telescopes such as the VLA in three main ways: 

\vspace*{-0.2cm}
\begin{itemize}
\item The $uv$ coverage is inherently continuous, and deconvolving is a refinement
rather than a necessity (see \S2.1.4). 
\item The periodic structure of the antenna generates low--level grating 
ring artifacts, which may, however, be reduced by subsequent processing 
(see \S2.2.1). 
\item While self--calibration is possible, it is not straightforward and 
requires some non--standard techniques (see \S2.1.4). 
\end{itemize}

Below, we describe the general properties of the MOST wide--field images 
used in the SUMSS, illustrated by examples from the first year of the survey. 
 
\subsubsection{Image size and resolution} 
As noted in Table 1, a 12-hour MOST wide--field observation centered at
declination $\delta$ yields an elliptical image of dimensions 
$163'\times 163'$\cd\ (RA $\times$ Dec), with a synthesised FWHM beam-size of
$43''\times 43''$\cd. The north--south extent of the field and the resolution
in the north-south direction both vary with declination as a direct result of
the foreshortening of the MOST  when observations are made away from the
meridian, as can be seen from Figure 2. Figure 2(a) shows a wide--field image
centered at declination $-80^\circ$; the synthesised beam
(43.0$''\times43.6''$)  and the field of view (2.73$^\circ\times2.77^\circ$)
are both essentially circular. Figure 2(b) shows an image centered at
declination $-33^\circ$, and here both the synthesized beam
(43.0$''\times77.3''$) and the overall field (2.73$^\circ\times4.91^\circ$) 
are extended north--south. 

\begin{figure}[tbp]
\fbox{These figures presented separately in jpg format}
  \caption{Two SUMSS images at different declinations: (a) SUMSS Field 80 at 
    declination $-80^\circ$, (b) SUMSS Field 2411 at declination
    $-33^\circ$.  Note the N--S elongation of the northern field, and
    the higher noise level at the northern and southern extremes of
    the field. 
    }
\end{figure} 

Less importantly, bandwidth and time--average smearing (Bridle \& Schwab 1989) 
can broaden the synthesised beam away from the field center. The effect is
small --- the point--spread function is broadened by up to 0.6$''$ 
in the direction tangential to the field center and by up to 3.3$''$ in the
radial direction. As a result peak flux densities may be reduced by up to 7\%,
but this is well--defined and can be corrected for in large--scale statistical
source analyses.  The integrated flux density is preserved. 

\subsubsection{Sensitivity}

\label{sec:noise}
The rms noise in individual MOST images is a combination of thermal noise and
source confusion from both the main beam and sidelobes of the MOST. The noise 
level varies with declination, position in the field, interference, weather
conditions and operational status of the telescope. 
Figure 3 shows how the rms noise at the field center varies with declination.
At declinations near $-70^\circ$ the rms noise at the center of a single
wide--field image is typically 0.8--1\,mJy beam$^{-1}$. 
\begin{figure}[tbp]
\vspace{0.5cm}
\centering\includegraphics[width=10cm]{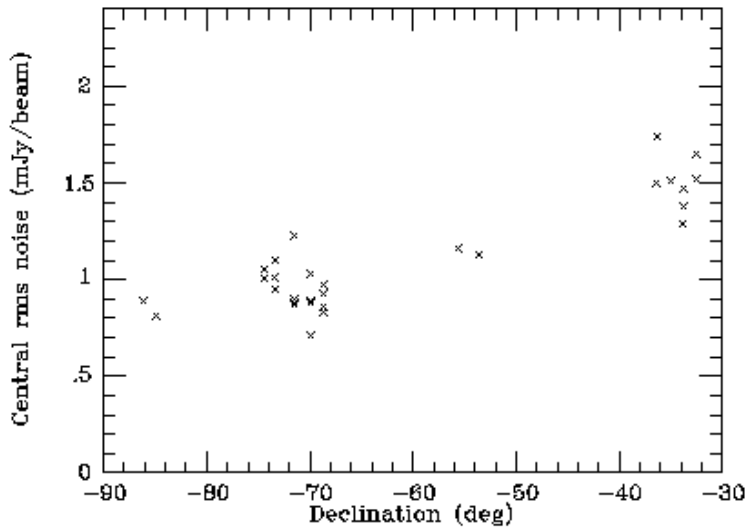}

\caption[]{Measurements of the rms noise near the field center, for data 
taken in the first month of the extragalactic survey for fields well away from 
the Galactic plane. The noise level is 
higher for the northern fields, as expected.  }
\label{fig:rms}
\end{figure}

The noise level in a MOST wide--field image varies across the field because 
of the complex antenna response of the MOST
in the east--west and north--south directions.  Examination of representative
fields, confirmed by approximate theoretical analysis,
shows that the noise increases with angular distance from the field
center. 
At 1.3$^\circ$ from the field center the rms noise is about 20\% higher 
than at the center, and increases more rapidly 
further out.  This is most obvious in northern fields such as the one 
in Figure 2(b), because of the `stretched' declination coverage. However, 
it is important to note that contours of constant noise level are 
approximately circular at all declinations.  

The final SUMSS data product will be a set of 4$^\circ \times 4^\circ$
mosaics, formed so as to recover uniform sensitivity and
dynamic range by appropriate overlapping of the observed individual fields. 

\subsubsection{Position and flux density calibration} 

The calibration of individual MOST images in both position and flux density is
based on observations of strong calibration sources (non--variable in flux
density, and of small angular size) which are made before and after each
12-hour synthesis observation\footnote{The desire for complete $uv$ coverage,
and the slow slew rate of the MOST, means that calibration sources cannot be
observed during a synthesis observation.}. The calibrators are chosen from
those listed by Campbell--Wilson~\& Hunstead (1994). 

Typically about 5 sources are observed using a 6-minute  observation known as a
SCAN.  Each SCAN yields three calibration factors: the `gain' parameter (in
Jy$^{-1}$), which defines the flux density scale; the `offset' parameter
(measured in arcseconds), which measures the mean pointing offset of the east
and west arms of the telescope in the meridian distance direction; and the
`phase' parameter (measured in degrees), which measures the phase error between
the two arms. Mean values of the SCAN parameters obtained before and after each
survey are used to calibrate positions and flux densities via the
Campbell--Wilson~\& Hunstead  (1994) list of calibrators. 

The flux density scale of our calibrator sources is based on the work of
Hunstead (1991). It relies on an interpolation between the scales of the
Molonglo Reference Catalogue (MRC) at 408\,MHz (the MRC scale is very 
close to that of Baars et al.\ 1977) and the Parkes Catalogue at
2700\,MHz, supplemented by absolute measurements at 843\,MHz.  

Tests on repeated fields show that for strong sources within 1$^\circ$ of the
field center, the rms error in flux density measurement is 5\%  relative to our
calibration sources. Within an individual field, the {\em relative}  flux
densities of unresolved strong sources can be measured with an rms error of 2\%
(slightly poorer near the edges of the field). 
  
For sources weaker than about 100\,mJy, thermal noise and confusion
increasingly affect the accuracy of flux density measurements, contributing
1--2 mJy to the uncertainty.  The final accuracy also depends on the chosen
source--fitting algorithm and the dynamic range of the observations.  A full
discussion of these topics will be given in the second paper (Cram et al.\
1998, in preparation). 

The absolute and relative accuracy of flux density measurements in the final
{\em mosaiced} images clearly  depends on the calibration  accuracy of the
fields comprising the mosaic. This is currently $\sim $ 5\%;  we are
investigating ways to improve the relative flux density calibration of the
constituent fields in a mosaic  {\it a posteriori\ }, for example by
reobserving a number of fields in the same night in partial synthesis mode to
tie their strongest sources together. 

For strong sources within 1$^{\circ}$ of the field center, the position
uncertainty is dominated by the accuracy to which the `offset' and `phase'
parameters can be determined from each night's SCAN calibrations.  This
produces typical (1$\sigma$) errors of about 1.0\,arcsec in right ascension and
1.0$\times\cd$\,arcsec in declination. 

For sources weaker than about 20\,mJy the final position measurements will be
limited by noise and confusion, and for resolved sources the choice of
source--fitting algorithm may also affect the measured positions.  Condon
(1997) gives a detailed discussion of how to derive error estimates for
elliptical Gaussians such as those used for radio--source fitting, and Condon
et al.\ (1998) discuss the estimated position errors for NVSS sources as a
function of flux density. 
We have made some simple tests which suggest that SUMSS and NVSS images have
comparable positional accuracy at similar flux densities. We used the AIPS task
VSAD to measure the positions and flux densities of sources in a 4.2\,deg$^2$
region of sky near  RA 22$^h$ 43, Dec $-39^\circ$ (J2000.0) . We then matched the
SUMSS sources with sources in the NVSS catalogue of the same region, and
tabulated the position difference for each source.  Within this region, there
were 101 matched sources with integrated SUMSS flux densities S$_{\rm 843}
>\,10$\,mJy.   For these sources, the mean offsets (NVSS--SUMSS) are as
follows: 

\noindent
\hspace*{2cm} $\langle \Delta$RA$ \rangle$ = $-0.176\pm0.204$\,arcsec \\
\hspace*{2cm} $\langle \Delta$Dec$ \rangle$ = $-0.129\pm0.314$\,arcsec 

We therefore find no evidence for any significant offset between the NVSS 
and SUMSS reference frames.  We will repeat this test with larger numbers of 
sources as the SUMSS survey progresses.  

Figures 4(a) and (b) show the measured position differences $\Delta$RA and
$\Delta$Dec between NVSS and SUMSS positions for 
(a) sources with flux
densities above 25\,mJy, and (b) weaker sources with flux densities in the
range 10--25\,mJy. These plots are analogous to Figures 28 and 29 of Condon et
al.\ (1998), where a similar comparison is made between positions measured by
NVSS  and FIRST for sources in common.  Following the practice of Condon et
al.\ (1998), we show 90\% confidence ellipses constructed under the following
assumptions: (i) SUMSS and NVSS position measurements of the same source have
the same uncertainty in RA, while the SUMSS uncertainty in Dec is equal to the
NVSS uncertainty multiplied by a factor \cd\ (=1.57 for the sources in Figure
4) to account for the stretching of the SUMSS beam in the declination
direction; and (ii) the semi--axes of the 90\% confidence ellipse are equal to
2.146 times the quadratic sum of the rms errors in the NVSS and SUMSS
positions. 

\begin{figure}[tbp]
\centering\subfigure[]{\includegraphics[width=7.5cm]{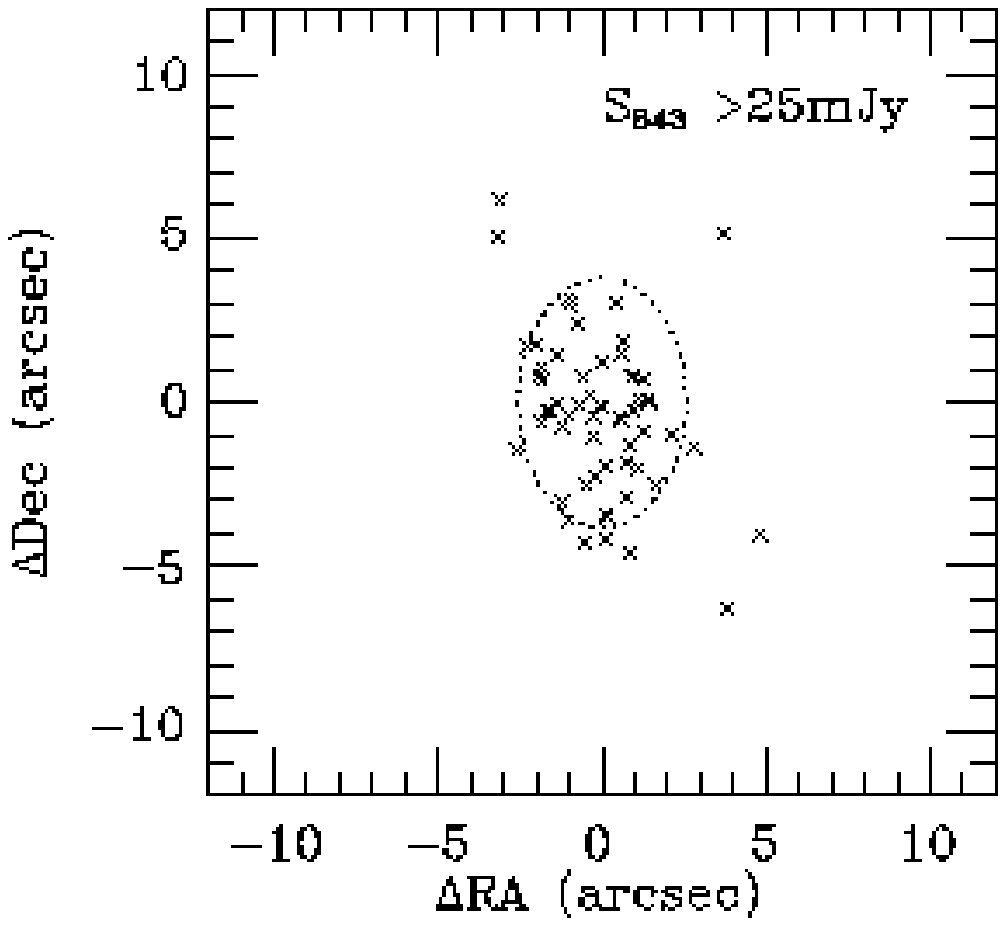}}
\hspace{1cm}
\subfigure[]{\includegraphics[width=7.5cm]{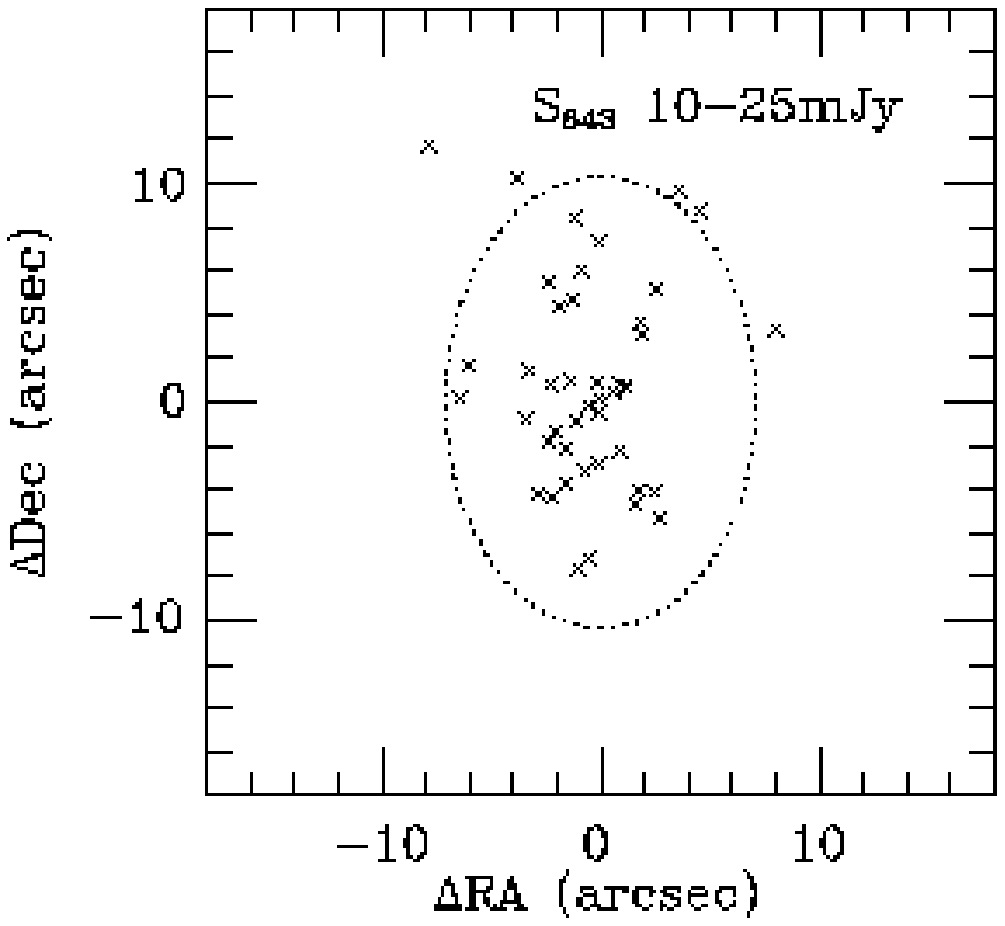}}
  \caption{(a) Position differences in RA and Dec between SUMSS and NVSS 
  measurements of common sources with total flux densities S$_{\rm 843} 
  > 25$\,mJy.  Dotted lines show the 90\% confidence error ellipse 
  calculated as described in the text.  
  (b) as (a), but for weaker sources with S$_{\rm 843}$ in the range 
  10--25\,mJy.  }
\end{figure}

Since most of the sources in Figures 4(a) and (b) lie within the 90\%
confidence limits described above, our assumption that NVSS and SUMSS sources
have similar position errors appears reasonable. We therefore estimate that
typical SUMSS position uncertainties in RA are $\sim5$\,arcsec at 5\,mJy,
2--3\,arcsec at 10\,mJy and $\sim1$\,arcsec above 20\,mJy.   Position
uncertainties in declination are
typically a factor \cd\ higher than in RA. 

\subsubsection{Dynamic range and the need for deconvolution} 
A strength of the MOST is the continuity of $uv$ coverage in the synthesized
images.   Apart from the gap due to the inter--arm spacing, all spacings are
represented out to the overall length of the instrument (1570\,m or
4400\,$\lambda$).  Figure 5 shows the radial profile of the $uv$ weighting of
the  MOST for a full (12--hour) synthesis. The weighting is essentially uniform
from 300\,$\lambda$ to 2500\,$\lambda$, then falls monotonically to zero at
4400\,$\lambda$. 
\begin{figure}[tbp]
\vspace{0.5cm}
\centering\includegraphics[height=7cm]{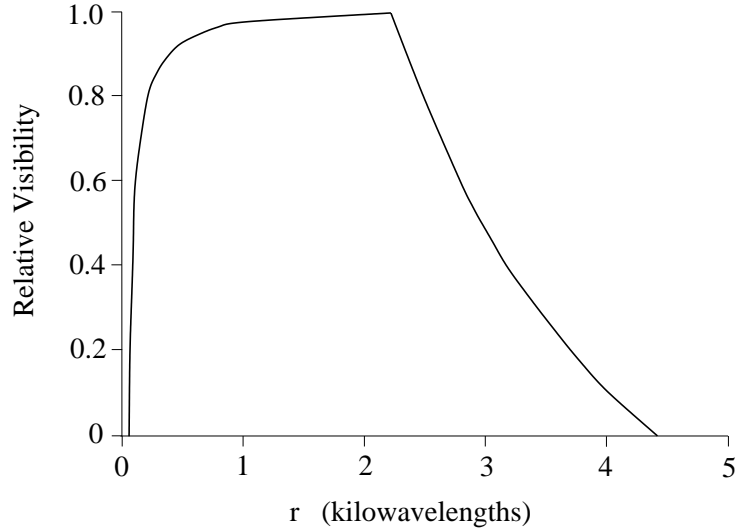}
\caption[]{A radial profile of the effective $uv$ weighting of the MOST's
  synthesised beam (after Durdin et al.\ 1984)}
\label{fig:uv}
\end{figure}

As a result, MOST raw images look rather like conventional optical telescope
images, with little need for deconvolving to make them visually acceptable or
astronomically useful.  This can be seen in Figure 6, which compares the
CLEANed and unCLEANed image of a complex radio galaxy. SUMSS images are
routinely CLEANed as part of the data reduction process, but the uniform
spatial frequency coverage in the raw images means that this process is
trouble--free, does not require subjective judgements to be made and does not
introduce spurious features. 
\begin{figure}[tbp]
\centering\includegraphics[height=20cm]{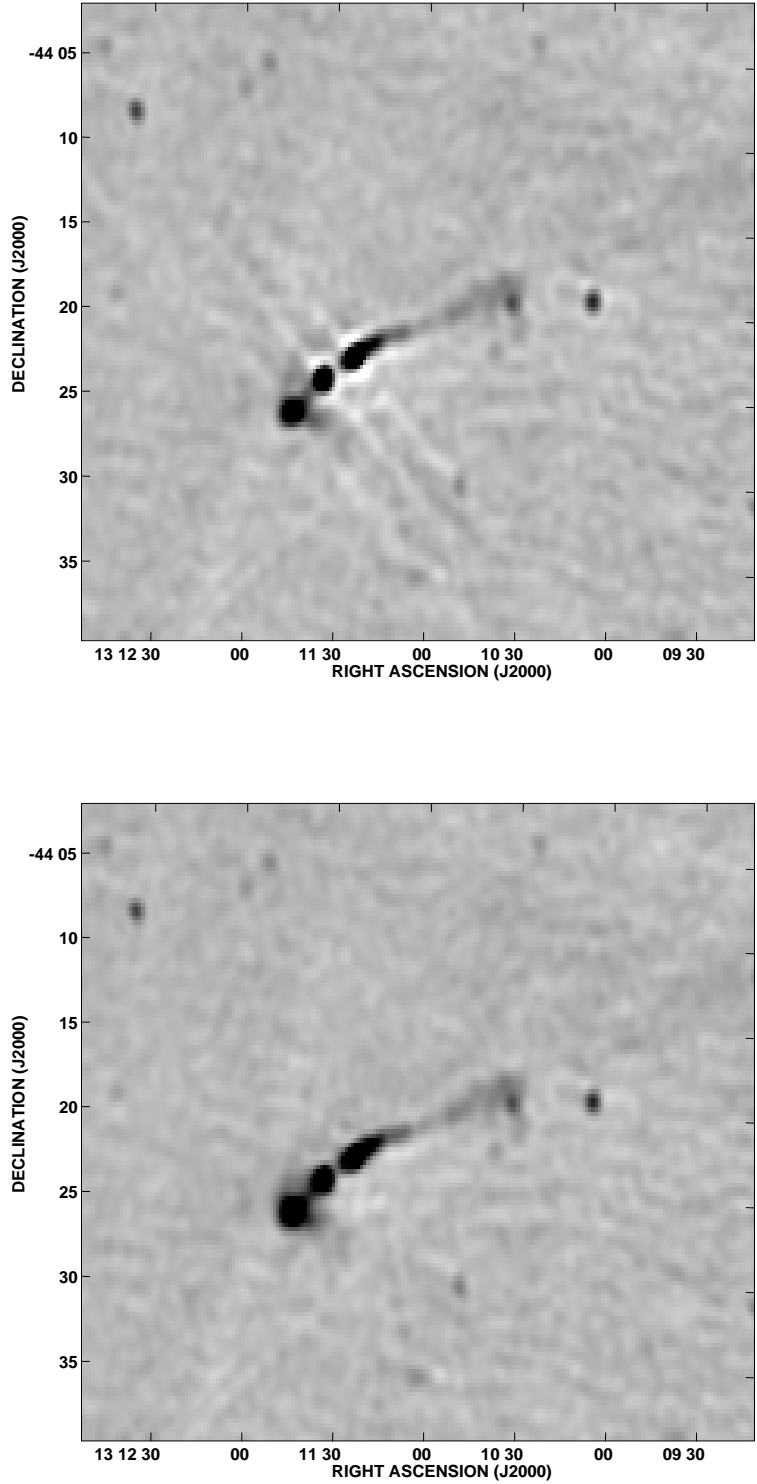}
 \caption{Effects of CLEAN on MOST images.  These two images show the radio 
   galaxy PKS\,1308$-$441 before (top) and after (bottom) CLEANing. Note that 
   that `dirty' MOST maps still have good cosmetic quality, even for complex 
   structures. } 
\end{figure}

In raw SUMSS images, gain variations and artifacts limit the effective dynamic 
range (i.e. the ratio of peak source flux density to peak artifact flux 
density) to about 100:1.  
Conventional self--calibration procedures involving
antenna--based corrections cannot be applied to MOST data since the signals
from element pairs are not recorded individually. For fields which include one
or more strong sources, an alternative technique known as `adaptive
deconvolution' (Cram \& Ye 1995) may be used: the three telescope fan--beam
parameters `gain', `offset' and `phase' (see \S2.1.3) are determined as a
function of time and applied to the raw data before the image is formed. In
some $70'$ fields (produced by the MOST in its original configuration), this
technique has improved the dynamic range by a factor of ten or more. Adaptive
deconvolution is used wherever possible to improve the dynamic range of SUMSS
images. 
 
\subsection{Image artifacts} 
MOST images are affected by a number of artifacts which, having a
characteristic appearance, can usually be distinguished from genuine
astronomical sources. In this section we describe the effect 
on the image of three main sources of artifacts: sidelobes, atmospheric
irregularities and interference. 

\subsubsection{Sidelobe response of the MOST } 
The only important sidelobe responses of the MOST are grating responses arising
from the periodic structure of the array.  In the images, the principal grating
lobe appears as an ellipse of semi-diameter  $4.6^\circ \times 4.6^\circ\
\,\cd$ centered on a strong source (which may be outside the synthesized area).
Although grating rings have a well--defined geometry, their amplitude is not
constant around the ring: it varies with position in a field, increasing with
distance from the center.  Where a ring passes through the exact center of a
field the amplitude is (ideally) zero.  A 1\,Jy unresolved source forms a
grating ring with an amplitude of up to about 10\,mJy beam$^{-1}$.

Because the amplitude of grating rings varies with distance from the field 
center in this way, grating rings in mosaiced images can change abruptly at
the boundaries of the fields comprising the mosaic. 
It is often possible to remove grating rings due to known unresolved sources,
though rings due to complex source structure are less tractable.  
 
\subsubsection{Atmospheric and ionospheric effects} 
Throughout all or part of a synthesis observation, the position of the comb of
fan beams may shift irregularly.  The shifts are typically $\sim1$\,arcsec, on
a timescale of minutes.  The causes of these shifts are currently under
investigation; they seem to be due in part to weather--related anomalies in the
distributed LO phase and in part to irregularities in ionospheric or
tropospheric refraction. The effect on images is to generate a pattern of
radial `spokes' about sources.   These features are weak, and are usually
visible above the noise only for unresolved sources stronger than about 100
mJy. 

Over a full 12-hour synthesis observation, the mean source displacement is zero
and there is negligible effect on the determination of the flux density or
source position.  Self--calibration is the most effective way to reduce or
eliminate the spokes. 

\subsubsection{Terrestrial interference }
The operating frequency of the MOST, 843\,MHz, is not in a protected radio
astronomy band. However, it was chosen in consultation with Telecom Australia
(now Telstra), and given some degree of local protection.  In 1996--97,
following joint field trials with the Australian Spectrum Management Agency
(Campbell--Wilson et al.\ 1997), local protection of the MOST band was
guaranteed until AD 2006. 

The observatory has experienced short bursts of interference from ground--based
transmitters, which were less frequent in 1997/98 than in previous years.  Such
interference is usually strong and easily recognized.  Since it is short--lived
and rare, the affected samples can be removed from the data before synthesis 
without greatly affecting the image quality.  Fortunately the off--axis gain of
the MOST is very low, except when an interfering source lies at the meridian
distance of the main beam or its grating responses. Also, the new MOST RF
and IF amplifiers have been designed to reject out-of-band
interference strongly (Campbell--Wilson et al.\ 1997). 

Radio telescopes worldwide are experiencing increasing levels of
radio--frequency interference, largely as a result of the communications
revolution in the last 5--10 years.  MOST is not immune to RFI, but at present
it does not seriously compromise our astronomical output. Despite
representations to the relevant spectrum management authorities our position is
not secure, and RFI is likely to become more severe in the future if usage of
the band increases without agreed principles for sharing. 

\subsubsection{Solar interference }
The Sun is the strongest naturally--occurring radio source in the sky, and it
can cause severe interference to MOST observations made in the daytime
(Campbell--Wilson et al.\ 1997).  Consequently, observations are scheduled as
far as possible at night, with the field center in transit within an hour or so
of midnight. When observations overlap daylight hours we, as far as possible,
plan them to minimize solar interference by  avoiding daytime observations at
declinations within 20$^\circ$ of the declination of the Sun. Solar
interference appears in images as parallel bands at predictable position
angles.   As with other forms of interference, the effects can often be reduced
to negligible levels  by suitable data processing. 

\section{The Sydney University Molonglo Sky Survey (SUMSS) and its goals }
We began the Sydney University Molonglo Sky Survey (SUMSS) in June 1997. In
designing the survey, we had to take into account both our science goals and
the observational constraints imposed by the MOST.  Here we describe how this
was done, and how we chose the survey parameters. 

\subsection{Survey design} 

The design of the survey is constrained by the structure of the MOST and by
our requirements for wide sky coverage, sensitivity and uniformity.  The main 
constraints are as follows: 

\vspace*{-0.2cm}
\begin{itemize} 
\item
Some time is needed each working day for routine maintenance of the telescope,
so we can schedule, on average, nine 12--hour observations each week.   As far
as possible, the observing program is designed so that fields observed at night
transit within an hour of midnight. 
\item
The MOST is an east--west array with limited meridian distance
coverage, and full synthesis is practicable only at declinations south of
$-30^\circ$ as discussed in \S2. Setting the northern limit of our survey at
$-30^\circ$ therefore ensures good--quality images while giving overlap with
the NVSS (Condon et al.\ 1998) in the declination range $-30^\circ$ to
$-40^\circ$. 
\item 
There are several reasons for aiming at a detection limit of about 5\,mJy 
for the survey. Figure 7 shows the predicted contribution of various radio
source populations as a function of flux density (Jackson \& Wall
1998).  
\begin{figure}[tbp]
  \begin{minipage}[b]{\textwidth}
 \centering  \includegraphics[height=12cm,angle=-90]{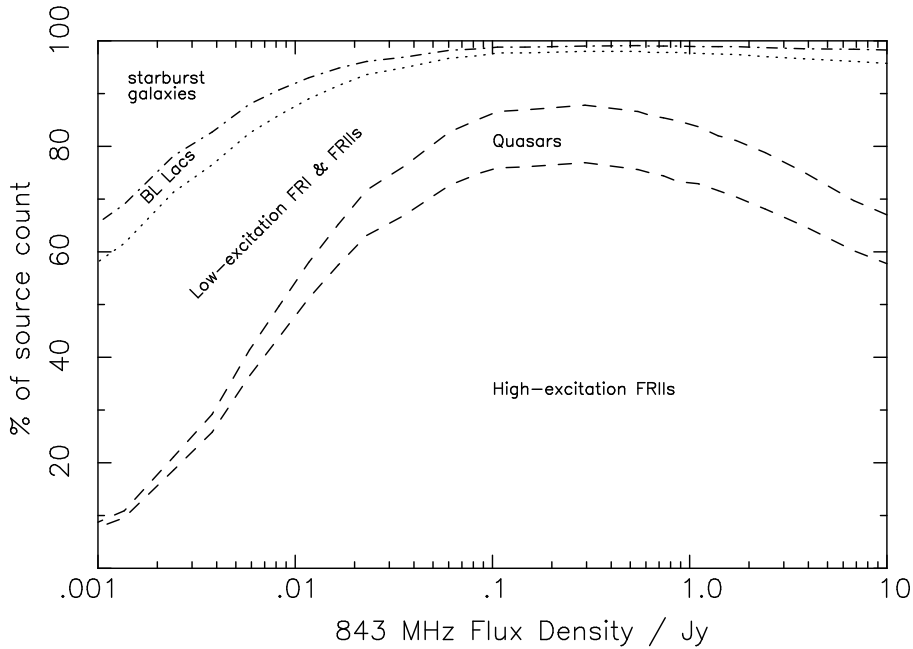}
   \vspace{0pt}
  \end{minipage}
  \caption{Predicted radio--source population mix at 843\,MHz, from 
    Jackson \& Wall (1998). } 
\end{figure} 
It predicts that by reaching a survey limit of 5--10\,mJy we will sample large
numbers of both AGN and star--forming radio populations, so that both
populations can be studied in detail. A 5\,mJy limit also yields a high enough
source density that studies of clustering and large--scale structure are
possible, and is well--matched to the typical (1\,$\sigma$) noise level of
1.0--1.5\,mJy for a single 12\,h synthesis observation with the MOST. 
\item 
We want to ensure (as far as possible) uniform sensitivity across a field and
from one field to another. What can be achieved  is limited by the variation of
rms noise level with declination (see Figure 3) and the center--to--edge
sensitivity variation of each field.  The best strategy is to make individual
observations on a grid of overlapped pointing centers which approximate to
hexagonal close packing, then mosaic the images to recover sensitivity in the
overlap regions.  The NVSS (Condon et al.\ 1998) uses a similar strategy. 
\item 
We need to cover at least 1000\,deg$^2$ per year to finish the survey in
reasonable time, and this in turn sets limits on the amount of overlap of
contiguous fields.   
\end{itemize} 

Each individual SUMSS field has a fully--synthesised elliptical area of
2.7$^\circ \times 2.7^\circ \times(\pi/4) \times$\,\cd\ (= 5.7\,\cd) square 
degrees at declination $\delta$. In defining our survey field centers, we have
assumed the `useful' field to be circular with diameter 2.6$^\circ$,
independent of declination.  This size was chosen because it corresponds to the
region in which the rms noise level at the edge is no more than $\sqrt{2}$
times the value at the field center, allowing us to restore uniform sensitivity
by incoherently combining the individual fields. We chose a pseudo--hexagonal
close packing of field centers, using a similar scheme to that described by
Condon et al.\ (1998) and adopted for the NVSS. (Figure 9 of the Condon et 
al.\ paper gives a useful schematic view of the field placing).  As for the
NVSS, our SUMSS field centers are arrayed on lines of constant declination,
with a spacing such that no point on the sky is more than 1.3$^\circ$ from a
field center.  With this field spacing, the average rate of progress (i.e. area
added to that already surveyed) is 4.4 square degrees per field, independent of
declination 

In the final grid, there are 2713 SUMSS field centers south of declination
$-30^\circ$. Table~2 lists the field centers and numbering
convention for individual survey fields.  Note that we have defined (and
numbered) `extragalactic' field centers for the whole sky south of declination
$-30^{\circ}$, even though we are not planning to use those within 10$^{\circ}$
of the galactic plane.   Our first priority is to complete the 1118
high--latitude fields with field centers at Galactic latitude $|b|>30^\circ$. 
\begin{table} 
\tablenum{2} 
\centering
{\small 
\begin{tabular}{ccccc|ccccc} \hline \hline
 Fields & $n$ & Dec & RA start & $\Delta$RA & Fields & $n$ & Dec & RA start & $\Delta$RA \\
          && (J2000.0) && & && (J2000.0) && \\
     1     &   1 &  $-$89 44 24   &   00 00 00    & --  &   698--745 &  48 &  $-$59 56 42   &   00 15 00    & 30 m \\
    2--7   &   6 &  $-$88 23 40   &   02 00 00    & 04h &   746--793 &  48 &  $-$58 46 48   &   00 00 00    & 30 m \\
    8--13  &   6 &  $-$87 23 24   &   00 00 00    & 04h &   794--841 &  48 &  $-$57 40 48   &   00 15 00    & 30 m \\
   14--25  &  12 &  $-$86 09 36   &   00 00 00    & 02h &   842--889 &  48 &  $-$56 35 24   &   00 00 00    & 30 m \\
   26--37  &  12 &  $-$84 55 12   &   01 00 00    & 02h &   890--985 &  96 &  $-$55 34 12   &   00 07 30    & 15 m \\ 
   38--49  &  12 &  $-$83 33 36   &   00 00 00    & 02h &   986--1081&  96 &  $-$53 35 24   &   00 00 00    & 15 m \\ 
   50--61  &  12 &  $-$82 25 48   &   01 00 00    & 02h &  1082--1177&  96 &  $-$51 41 24   &   00 07 30    & 15 m \\ 
   61--73  &  12 &  $-$81 20 24   &   00 00 00    & 02h &  1178--1273&  96 &  $-$49 53 24   &   00 00 00    & 15 m \\ 
   74--97  &  24 &  $-$80 37 48   &   00 30 00    & 01h &  1274--1369&  96 &  $-$48 13 12   &   00 07 30    & 15 m \\ 
   98--121 &  24 &  $-$79 18 00   &   00 00 00    & 01h &  1370--1465&  96 &  $-$46 42 36   &   00 00 00    & 15 m \\ 
  122--145 &  24 &  $-$78 01 48   &   00 30 00    & 01h &  1466--1561&  96 &  $-$45 24 24   &   00 07 30    & 15 m \\ 
  146--169 &  24 &  $-$76 45 00   &   00 00 00    & 01h &  1562--1657&  96 &  $-$44 06 36   &   00 00 00    & 15 m \\ 
  170--193 &  24 &  $-$75 34 12   &   00 30 00    & 01h &  1658--1753&  96 &  $-$42 48 36   &   00 07 30    & 15 m \\ 
  194--217 &  24 &  $-$74 25 12   &   00 00 00    & 01h &  1754--1849&  96 &  $-$41 30 36   &   00 00 00    & 15 m \\ 
  218--265 &  48 &  $-$73 23 24   &   00 15 00    & 30 m & 1850--1945&  96 &  $-$40 13 12   &   00 07 30    & 15 m \\ 
  266--313 &  48 &  $-$71 31 48   &   00 00 00    & 30 m & 1946--2041&  96 &  $-$38 55 48   &   00 00 00    & 15 m \\ 
  314--361 &  48 &  $-$69 58 48   &   00 15 00    & 30 m & 2042--2137&  96 &  $-$37 38 24   &   00 07 30    & 15 m \\ 
  362--409 &  48 &  $-$68 40 12   &   00 00 00    & 30 m & 2138--2233&  96 &  $-$36 21 00   &   00 00 00    & 15 m \\ 
  410--457 &  48 &  $-$67 22 48   &   00 15 00    & 30 m & 2234--2329&  96 &  $-$35 04 12   &   00 07 30    & 15 m \\ 
  458--505 &  48 &  $-$66 05 24   &   00 00 00    & 30 m & 2330--2425&  96 &  $-$33 48 00   &   00 00 00    & 15 m \\ 
  506--553 &  48 &  $-$64 49 12   &   00 15 00    & 30 m & 2426--2521&  96 &  $-$32 31 12   &   00 07 30    & 15 m \\ 
  554--601 &  48 &  $-$63 33 00   &   00 00 00    & 30 m & 2522--2617&  96 &  $-$31 15 36   &   00 00 00    & 15 m \\ 
  602--649 &  48 &  $-$62 19 48   &   00 15 00    & 30 m & 2618--2713&  96 &  $-$29 59 24   &   00 07 30    & 15 m \\ 
  650--697 &  48 &  $-$61 06 36   &   00 00 00    & 30 m &           &     &              &               &      \\ 
\hline
\end{tabular}
}
\caption{Field centres for individual SUMSS observations } 
\end{table}

Table 3 compares the SUMSS survey with the three northern hemisphere 
radio imaging surveys now in progress or recently completed.  At the time 
of writing (November 1998), the SUMSS survey is about 15\% complete.  
\begin{table}
\tablenum{3}
\centering
\begin{tabular}{lcccc}
\hline
 & \multicolumn{1}{c}{ FIRST }  & \multicolumn{1}{c}{ NVSS } &  
   \multicolumn{1}{c}{ \bf SUMSS } & \multicolumn{1}{c}{ WENSS } \\
\hline
Frequency (MHz)  & 1400    & 1400        & 843      & 325      \\
Area (deg$^2$) & 10,000  & 33,700      & 8,000    & 10,100   \\ 
Resolution   & 5$''$     &  45$''$     & 43$''$   & 54$''$   \\ 
Detection limit&  1\,mJy & 2.5\,mJy    & 5\,mJy & 15\,mJy  \\
Coverage     & NGP & $\delta>-40^\circ$ & $\delta<-30^\circ$ & 
$\delta>+30^\circ$ \\
Sources/deg$^2$ & 90     &   60        & 37       & 21      \\
\hline
\end{tabular}
\caption{Comparison of SUMSS with northern radio imaging surveys}
\end{table}

\subsection{Data release} 

We intend to make the SUMSS data publicly available as the survey proceeds. 
The final data products will be a set of $4^\circ\times4^\circ$ mosaiced 
images (using the same mosaic centers as the NVSS but extending further 
south) and a source catalogue.  Further information (including regular 
updates on the progress of the survey) can be found at our 
survey web site, http://www.physics.usyd.edu.au/astrop/SUMSS.  

\section{Expected results of the survey } 

We now give a brief outline of the kinds of scientific problems which can 
be tackled with SUMSS data, either alone or in conjunction with data 
at other wavelengths. 

\subsection{Radio source counts} 
Radio source counts at 843\,MHz have already been measured over areas of 
several square degrees by Subrahmanya and Mills (1986).  They used repeated 
observations of several fields to improve the sensitivity, and covered 
a total area of 0.4 deg$^2$ for sources in the range 1--2\,mJy, 2.3 deg$^2$ 
for sources in the range 2--7\,mJy and 20 deg$^2$ for sources in the range 
7--224\,mJy.  Their data were further analysed by Large (1990), who derived 
and tabulated integral source counts for the Subrahmanya \& Mills survey 
area.  Since Large's original (1990) paper had very limited circulation, 
the integral source counts are reproduced here as Table~4. 
\begin{table}
\tablenum{4}
\centering
\begin{tabular}{rrr}
\hline
\multicolumn{1}{c}{S$_{843}$}  & \multicolumn{2}{c}{ N($\geq$S)} \\
\multicolumn{1}{c}{(mJy)} & \multicolumn{1}{c}{per} & \multicolumn{1}{c}{per} \\
          & \multicolumn{1}{c}{steradian} &\multicolumn{1}{c}{square degree} \\
\hline
  1       & 432,000    &  132 \\
  2       & 249 000    &   76 \\ 
  3       & 180 000    &   55 \\
  5       & 120 000    &   37 \\
 10       &  68 200    &   21 \\
 20       &  38 300    &   12 \\
 50       &  17 100    &    5.2 \\
 100      &   8 620    &    2.6 \\ 
 200      &   3 720    &    1.1 \\
\hline
\end{tabular}
\caption{Integral source counts at 843\,MHz, from Large (1990)}
\end{table}

While the SUMSS images will not go deeper than the earlier MOST source count
studies, they will cover a much larger area, allowing us to compare the source
counts in several areas.  We would expect {\it a priori}\ to find a similar
result wherever we look, but this will not be the case if the weakest sources
(1--10 mJy) are clustered. 

\subsection{Radio spectral index measurements} 
The region to be observed with MOST partly overlaps the northern NVSS survey
(Condon et al.\ 1998), which  extends down
to $-40^\circ$ declination (see Table 3).  The area of overlap is 800~deg$^2$.
The NVSS images have similar angular resolution and sensitivity to the SUMSS
(see Table 3), but are at a higher frequency (1400\,MHz compared with
843\,MHz). 

Figure 8 shows a comparison between a 5 deg$^2$ SUMSS mosaic and the
corresponding region from NVSS.  In general the images are
similar, though as noted earlier SUMSS has superior $uv$ coverage and will
therefore record extended emission regions with greater fidelity. 
\begin{figure}[tbp]
\fbox{This figure presented separately in jpg format}
  \caption{Comparison of NVSS and SUMSS mosaics for the ESO imaging 
survey (EIS) region.} 
\end{figure}

Figure 9 shows the spectral index distribution between 843\,MHz and 1.4\,GHz
for a sample of 195 sources common to both surveys.  
\begin{figure}[tbp]
\includegraphics[height=9cm]{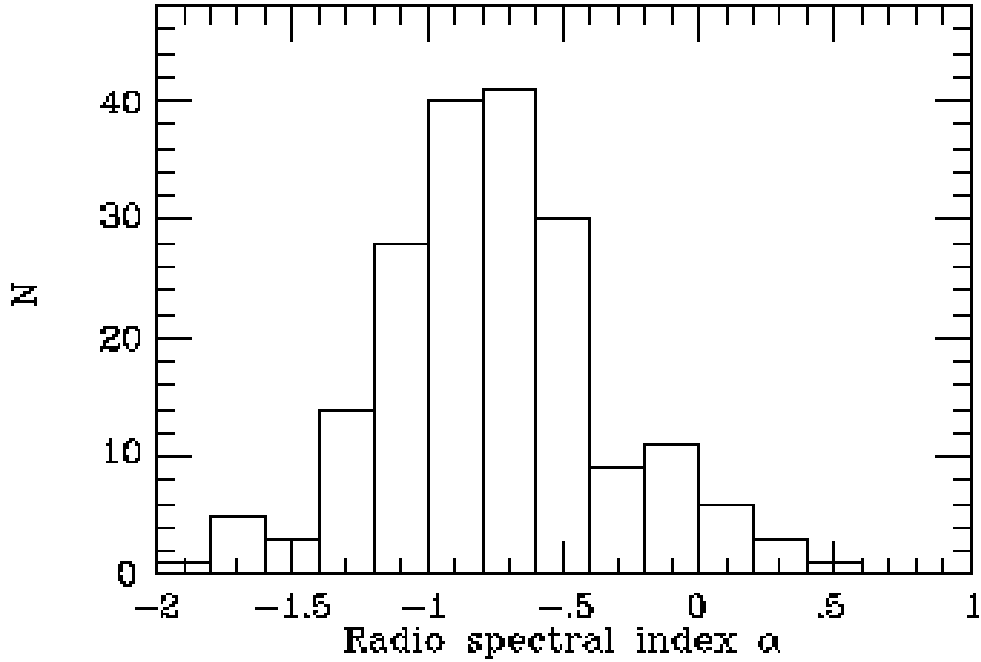}
 \caption{Radio spectral index distribution between 843\,MHz and 1.4\,GHz 
   for 195 sources with S$_{\rm 843} >\,10$\,mJy observed by both NVSS and 
   SUMSS } 
\end{figure} 
The distribution has a
median spectral index $\alpha=-0.8$ (for $S \propto \nu^{-\alpha}$).  About 
20\% of the sources have $\alpha < -1.3$ or $> -0.3$.  Although
care is needed in interpreting radio spectra obtained over a relatively short
frequency baseline, it appears that we will be able to identify rare, but
astrophysically interesting objects such as faint ultra-steep-spectrum (USS)
sources which are likely to be galaxies at very high redshift (R\"ottgering
1993; Spinrad 1994). 

SUMSS will also yield spectral index information for several thousand strong
southern radio sources through cross--identification with the PMN source
catalogue (Griffith \& Wright 1993), which lists sources stronger than about
50\,mJy at 5\,GHz. Figure 10 shows a 17\,deg$^2$ region of a SUMSS mosaic of
the southern sky. 

Open circles mark the positions of PMN sources from an online
version of the catalogue supplied by Dr Alan Wright.  The higher sensitivity of
the SUMSS data can clearly be seen.  There are 57 PMN sources in this region,
and 9 of these are clear doubles in the SUMSS images.  The positional agreement
with PMN is generally good, though two PMN sources have no SUMSS counterpart
and another close pair of PMN sources corresponds to a single SUMSS source. 
\begin{figure}[tbp]
\fbox{This figure presented separately in jpg format}
  \caption{17 deg$^2$ SUMSS mosaic of a region in the southern sky.  
    Open circles show positions of PMN sources detected at 5\,GHz. }
\end{figure} 
Although the surface density of PMN sources is low compared to SUMSS, we expect
to have about 15,000 PMN sources in our survey region. With such a large sample
we can look for differences in the spectral index distribution as a function of
flux density, optical identification, etc. We will also measure more accurate
positions for the PMN sources, so that optical identification of the entire PMN
sample can be attempted for the first time. 

\subsection{The two--point correlation function and source clustering } 
SUMSS will cover much of the southern sky to a depth of $\sim$\,37  sources
deg$^{-2}$.  Such sensitivity --- corresponding to a 5$\sigma$ flux density
limit of 5\,mJy at 843 MHz --- is unprecedented in the south. Earlier
southern surveys had source densities of 0.5 deg$^{-2}$ for the 408~MHz
Molonglo Reference Catalogue (MRC; Large et al.\ 1991) and 2--3 deg$^{-2}$ for
the PMN 5~GHz survey (Griffith \& Wright 1993). 

The apparent isotropy of the radio sky reported from analysis of strong--source
surveys arises mainly through sparse sampling of the source population: to
detect anisotropies, a survey must reach a level faint enough for large
structures to contribute more than one source to the survey. There is now
convincing evidence for large--scale structure in the spatial distribution of
faint radio sources (Wall et al.\ 1993, Kooiman et al.\ 1995). Cress et al.\
(1996) have measured the angular two--point correlation function for radio
sources above 1\,mJy in a 1550 square degree region of sky covered by the FIRST
survey, and show that faint radio sources are clustered on scales of 0.2 to 2
degrees. 

Several clustering studies are possible with SUMSS data; for example we can
measure the two--point correlation function over large areas of sky, provided
that our data are sufficiently uniform in quality. 
The wide--field MOST images provide information on structures $\sim$100 Mpc or
larger, i.e. intermediate between the $\sim$1\,Gpc scales sampled by the COBE
microwave background fluctuations and the 10--100\,Mpc scales probed by optical
and IRAS galaxy-clustering studies (eg. Geller \& Huchra 1989; Maddox et al.\
1990; Ramella et al.\ 1992; Benn \& Wall 1995). 

Without redshift measurements, however, radio--source surveys probe only the
projected distribution of galaxies (radio flux densities are essentially
uncorrelated with redshift, because of the broadness of the radio luminosity
function). Dunlop \& Peacock (1990) show that a combination of source counts
with incomplete identifications and measured redshifts can be used to estimate
the distribution of radio sources with redshift for different flux density
limits. The first detection of the 3D clustering of radio sources was made by
Peacock \& Nicholson (1991), based on a redshift survey of 329 galaxies with
0.01 $<$ z $<$ 0.1 and a 1400\,MHz flux density above 500\,mJy. 

If we can determine the statistical properties of the redshift distribution of
SUMSS sources as a function of flux density, it would be possible to study
radio source clustering in three dimensions over a wide redshift range.  This
would be an ambitious project, requiring considerable follow--up time on
optical telescopes. 

\subsection{Identification of SUMSS radio sources at other wavelengths } 
Cross--comparison with optical catalogues will allow us to determine which
members of known classes of optical objects (for example nearby galaxies, or
optically-selected QSOs) are detected as radio sources.  Evolution in their
radio properties with redshift is an important cosmological indicator and 
there are special statistical tools --- such as survival analysis (Isobe et
al.\ 1986) --- to deal with low detection rates. 

\subsubsection{Optical identification from Digital Sky survey (DSS) images } 
The COSMOS database (Yentis et al.\ 1992) is a powerful tool for automatic
identification of radio sources (Unewisse et al.\ 1993), and MOST positions are
accurate enough for unambiguous identification of sources stronger than
about 10\,mJy, where the position errors are $\lesssim 1-2''$.  For the
faintest sources ($\sim$5 mJy) the error climbs to $\sim5''$ (see \S2.1.3), 
and tests using real data and randomly--offset radio positions suggest 
that the completeness and reliability of position--based optical
identifications falls to about 50--60\%. 

Experience with current MOST images shows that 25--30\% of sources above
5\,mJy have an optical counterpart recorded by COSMOS (with the proviso
noted above).  For a 5\,mJy detection limit, we therefore expect to detect 
at least 40 sources per field with COSMOS counterparts.  

Current COSMOS data cover the blue (B$_{\rm J}$) band, but two--colour (B,R)
data from the Cambridge APM survey are expected to be available soon, and 
will give us a powerful way of identifying rare classes of objects. For
example, the optical counterparts of the most distant (z $>$ 4) quasars are
extremely red (B$-$R $>$3; McMahon \& Irwin 1991).  It would be
straightforward to produce a list of very red optical objects 
associated with radio sources for spectroscopic follow--up, in the 
expectation that many would be high--redshift radio--loud quasars. 

\subsubsection{Cross--identification with other catalogues} 
Cross--comparison with optical catalogues will allow us to determine which
members of known classes of optical objects are detected as radio sources.
Catalogues we plan to use include: 

\noindent
$\bullet$ The Third Reference Catalogue of Bright Galaxies (RC3;de Vaucouleurs
et al.\ 1991). The detection rate of RC3 galaxies (mostly spirals) is
$\sim$30\% in the the few SUMSS fields for which a comparison has so far been
made; further observations are required to establish better statistics. About
2000 RC3 galaxies fall within the area to be surveyed by SUMSS. 

\noindent
$\bullet$ The 2dF galaxy redshift survey (Colless 1998). The 2dF survey will
measure redshifts for a complete sample of 250,000 southern galaxies brighter
than B magnitude 19.5 over an area of 1700 deg$^2$. About 2\% of these galaxies
are detected as radio sources by SUMSS and/or NVSS. By combining SUMSS data
with the 2dF galaxy redshift survey results, it will be possible to measure the
843\,MHz flux density or upper limit for a sample of about 75,000 `normal'
galaxies in the redshift range $ 0\leq z \leq 0.3$. 

\noindent
$\bullet$ The Abell--Corwin--Olowin (ACO) catalogue of optically selected
rich clusters of galaxies (Abell et al.\ 1989). The catalogue extends to a
nominal redshift of 0.2, and lists hundreds of clusters south of 
declination $-30^{\circ}$.

\noindent
$\bullet$ An X--ray flux--limited sample of Abell clusters of galaxies compiled
from ROSAT  (Ebeling et al.\ 1996). This sample of very X--ray luminous
clusters, which is essentially X--ray selected and not affected by optical
extinction, contains dozens of clusters south of declination $-30^{\circ}$. 

\subsubsection{Cross--identification with IRAS sources} 
An important tool for identifying star-forming galaxies {\it without\/} the
need for optical spectroscopy makes use of the remarkably tight correlation
between radio and far-infrared (FIR) luminosities (e.g. Wunderlich et al.\ 
1987, Condon et al.\  1991).  
For spiral galaxies, we find $S_{60\mu{\rm m}} \sim 100\, S_{\rm 843\,MHz}$, 
so MOST should detect nearly all galaxies above the IRAS Faint Source Catalog
limit of 0.28\,Jy at $60\mu{\rm m}$. The surface density of star--forming
galaxies detected in this way is about 1 per square degree, or 4--5 galaxies 
per SUMSS field.  

\section{Summary } 
The Molonglo Observatory Synthesis Telescope (MOST) has recently been upgraded
to widen the field of view to 5 square degrees.  This has enabled us to begin
a deep radio imaging survey of the southern sky, with sensitivity and
resolution similar to those attained by the northern NRAO VLA Sky Survey
(NVSS).  The reduced survey data will be made publicly available as the
survey proceeds, and can be used to tackle a wide variety of astronomical
problems.  

\acknowledgments

The Molonglo Observatory Site Manager, Duncan Campbell--Wilson, and the 
technical staff Jeff Webb and Michael White have made invaluable
contributions to the installation and commissioning of the new wide--field 
hardware.  They, together with our recent appointees at the Observatory,
John Van Beekhuizen and Nancye Westworth have continuing responsibility
for the telescope maintenance and for running the SUMSS observational
program.  We warmly thank them for their dedication to these demanding tasks.
We also thank Ralph Davison, who has been responsible for much of the design
and fabrication of the new electronic components on which SUMSS depends, and
Barbara Piestrynski who maintains the data archive.  We gratefully 
acknowledge the contributions made by Fred Peterson, the Physics workshop
and, last but not least, the academic staff, students and visitors who have 
been involved with the upgrade of the telescope for many years.   The MOST
is operated with the support of the Australian Research Council and the 
Science Foundation for Physics within the University of Sydney.
  
We also thank the anonymous referee for several useful comments and 
corrections.

\newpage


\noindent
{\large \bf APPENDIX\\}

\noindent
When the MOST was developed from the 408 MHz Mills Cross, the field size 
was limited by the available funding and by the configuration of the Cross.
With the growing interest in exploiting beam--forming techniques in the next
generation of radio telescopes (e.g. for the Square Kilometer Array), it is 
of interest to report here the techniques used to more than quadruple the 
original field of view of the instrument.

The MOST was formed by converting the east--west arm of the Mills Cross into a
steerable phased array operating at 843 MHz.  For this purpose the arm was
divided into 88  bays, each $50\, \lambda$ long. The field of the MOST was
essentially the  ``primary'' beam of an individual bay.   In the {\em tilt}
direction the bay beamwidth (FWHM) was $\sim 2.3^{\circ} $, related to the
physical width of the reflector.   In the {\em meridian distance} (MD)
direction it was $1.2 ^{\circ} \sec $ MD, set by the $50\, \lambda$ length of a
bay. The bay length was thus a crucial parameter in the original design of the
MOST, determining the field size and speed as an imaging telescope as well as
the overall cost and complexity of the conversion. 

A second funding--limited factor was the size of the correlator or
multibeamer.     The analog multibeaming system forms a {\em block} of 64
hard--wired fan beams, having a profile appropriate earth--rotation image
synthesis. These  beams are equally spaced at near the Nyquist interval of
$22''\sec$ MD.\footnote{The beams are interlaced in time with an 
11\arcsec\ offset
to simplify interpolation during data processing.} The width of this block
of beams limits the field size realized by the MOST (in its basic mode of
operation) to $23'\times23'$\cd (R.A. $\times$ dec.).  Sources lying
outside this field, but within the primary beam of the bays, are partially
synthesized. 

In the  original design of the MOST, provision was made to increase the
fully synthesized field size by a time--sharing technique in which the comb
of beams was offset  $ \pm 23'\sec $ MD  from the field center in a 24 s
cycle. This technique increases the field to $70'\times70'$\cd, a nine--fold
increase in sky area.  It also results in a loss of sensitivity and the
augmentation of grating--ring sidelobes. 

In 1989, each bay was divided in two, doubling the number of independent 
antenna elements in the MOST from 88 to 176.  At the same time a simple 
phase switching system was
installed which allowed the beam of a bay to be
offset in synchrony with the offset of the comb of fan beams.   The result
was a marked improvement in the quality of the $70'\times70'$\cd\ images.
(Amy \& Large 1990). 

The success of the 1989 technique encouraged us to investigate methods for
increasing the size of the fully synthesized field to make full use of the
$\sim 2.3^{\circ}$ width of the primary beam in the tilt coordinate. An
increased field size could be achieved only by increasing once more the
number of independent elements in the array, enabling the formation of more
independent fan beams. For practical and economic reasons we decided to
increase the number of beams by extending the time--sharing principle,
rather than by constructing a larger multibeamer or correlator. Four new
low noise preamplifiers (LNA) were installed in each bay (ie one LNA for
each 6.25 $\lambda$ length of the arm), together with computer--controlled
phase shifters.  The configuration is shown in Figure 11.  
\begin{figure}[tbp]
\centering\includegraphics[height=10cm]{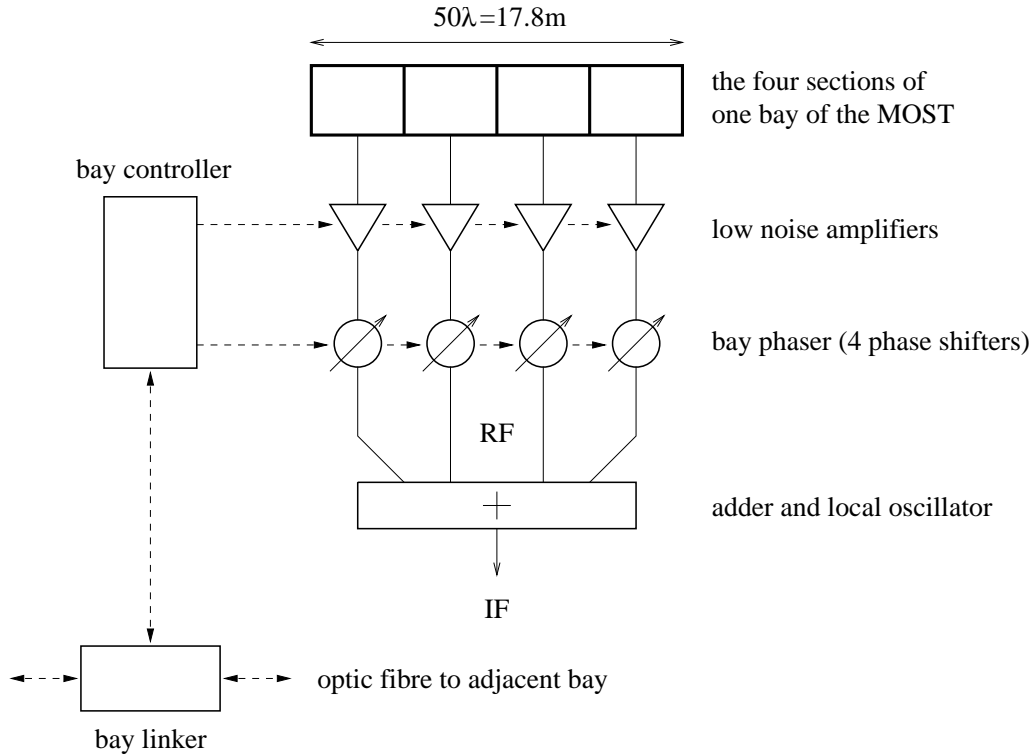}
  \caption{Hardware implementation of the Wide Field system.   The low noise
   amplifiers, bay phaser, bay controller and bay linker are the components
   of the hardware installed at each of the 88 bays.   The radio frequency
   (RF) and intermediate--frequency (IF) signal paths are shown as solid 
   lines.   Dashed lines indicate the control signal paths.   Omitted for
   clarity are gain compensation circuits in the bay phaser and ancilliary
   control inputs and outputs on the bay controller.}
\end{figure}
The purpose of
the phase shifters is to apply a (rapidly changeable) four--step phase
gradient to the bay, thus offsetting its beam.  Shifting the bay--beams
in this way is analagous to the short antenna slews used in 
close--packed mosaicing observations with the Australia Telescope
Compact array and the VLA. 
Electronic control of the phase shifters allows the field to be widened  by
cyclically offsetting  the primary beam of the bays to each of N meridian
distances from the nominal (tracked) field center.   These offsets of the
primary beam are synchronized with corresponding offsets of the comb of
fan beams. The technique can be understood with reference to
Figure 12.  Figure~12~(a) shows the beams at
meridian transit superimposed on the field (shaded circle) to be imaged by
time--sharing. The grating lobes which inevitably arise from the periodic
bay structure of the MOST are also shown.  
\begin{figure}[tbp]
  \begin{minipage}[b]{\textwidth}
\includegraphics[angle=-90,width=\textwidth]{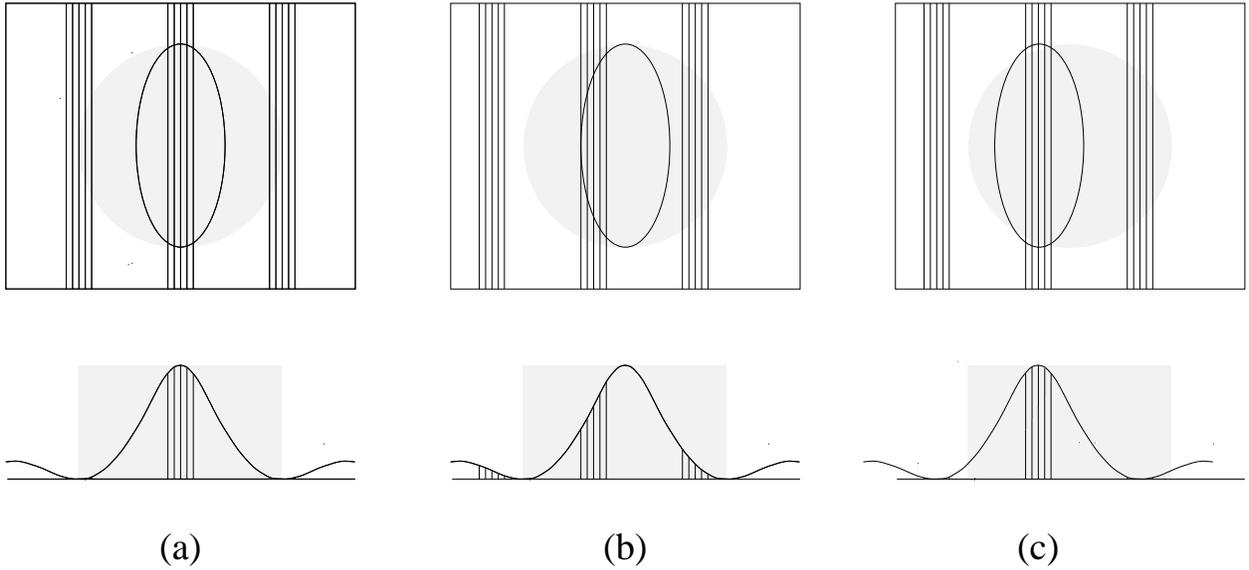}
  \end{minipage}
  \caption{Widening the field of view of the
    MOST by time sharing. A representation of the MOST beams at the
    mid point of a 12 hour synthesis with the field to be synthesized
    shown as a shaded circle.  In (a) the fan beams and the bay
    primary beam are located centrally on the field.  The grating
    lobes (see text) are largely suppressed as they lie about the
    nulls of the bay primary beam, as seen in the profile.  Fig. (b)
    shows the two problems that arise if the time-sharing is attempted
    by simply offsetting the fan beams: the fan beam gains are reduced
    and the grating responses are enhanced as they no longer lie on
    primary beam nulls.  These problems are overcome if the primary
    beams are offset with the fan beams.  Fig. (c) shows one of the 7
    offsets used to synthesize a wide field.}
\end{figure}
As can be seen in the profile,
the fan beams are centered on the peak of the bay beam, while the 
first--order grating responses due to the bay periodicity are 
small as they lie at
the nulls of the bay beam. Figure~12~(b) illustrates two
problems which arise with the time--sharing technique when the comb of
beams alone is offset: 

\begin{itemize}
\item the gain of the principal fan beams is reduced;
\item the unwanted grating responses are increased as they are shifted from the
  nulls of the bay beam.
\end{itemize}

Both these problems are solved by offsetting the bay beams synchronously
with the fan beams, as shown in Figure~12~(c).  The
implementation of this facility to offset the bay beams rapidly and
simultaneously with the fan beams is the essence of the upgrade which makes
the survey described in this paper practicable. 

Before 1997, the usual field size of MOST observations was
$70'\times70'$\cd, the comb of beams being cycled through three positions.
In the wide--field mode used for the SUMSS, the beams are (usually) cycled
through 7 positions.   The observing time spent on each position is thus
reduced by a factor of 7/3 and the expected signal--to--noise reduced by a
factor of $\sqrt{7/3}$ ($\sim 1.5$).  The reduction in signal--to--noise has
proved to be quite small, as the new preamplifiers have a markedly lower
input noise.  The 7/3 increase in the meridian distance range covered by
the time sharing increases the fully synthesized field area by a factor of
$(7/3)^{2}$ ($\sim 5.4$).  In practice the increase of usable field area is
somewhat less than the factor of 5.4, since the signal--to--noise
deteriorates towards the extreme edge of the field as discussed in the
main text. 

The field area imaged  by the MOST  is now limited by the width of the
primary (bay) beam in the tilt direction.  Consequently, no further
increases in the field size are possible if full synthesis of the field is
to be completed in one 12 hour observation. However, in principle the data
acquisition rate of the MOST could be increased by further increases in the
number of independent antenna elements and fan beams.  The speed of the
instrument for making imaging surveys of large areas would increase pro
rata, but full synthesis  would depend on combining data from many days'
observation. 

%
%


\end{document}